# Simulation as a sustainable trading zone: Aiming at intergenerational justice[1]

V. Pronskikh[2]

Fermi National Accelerator Laboratory, Batavia, IL 60510-5011, USA

Abstract

The paper, drawing on the example of simulation codes used in nuclear physics and high-energy physics, seeks to highlight the ethical implications of discontinuing support for simulation codes and the loss of knowledge embodied in them. Predicated on the concept of trading zones and actor network models, the paper addresses the problem of extinction of simulation codes and attempts to understand their evolution and development within those frameworks. We show that simulation codes of closed type develop to the level of creoles, becoming local languages and standards of scientific centers and disappearing as their few main developers leave, whereas codes of open types become universal languages, imposing problem-solving patterns on the entire community and crowding out other codes. The paper suggests that because of simulations' reliance on tacit knowledge, practices entrenched in codes cannot be exhaustively explicated or transmitted through writing alone; on the contrary, the life cycle of a simulation code is determined by the life cycle of its trading zone. We examine the extent to which both of these phenomena pose a risk to the preservation of knowledge. Bearing upon intergenerational ethics, we draw analogies between the pure intergenerational problem (PIP) and the problem of preserving the knowledge implemented in simulation codes and transmitting it to future generations. We argue that for the complete transfer of knowledge, it is necessary to develop and maintain inhabitability and sustainability of simulation trading zones in a controllable way, at least until the demand for these codes is warranted to cease in the future.

---

1. **Introduction**

Computer simulations and computer models are phenomena that have emerged and developed rapidly over the past decades, becoming an integral part of not only all branches of science and technology but even the humanities and social sciences. Numerous methodological and epistemological problems have come to the fore in current discussions of models and simulations, and have been the subject of philosophical analysis in a number of substantial studies. (1–9) Philosophers have long focused on the similarities and dissimilarities among modeling, simulation, and experimentation; the epistemological novelty of simulations; the problem of materiality in simulations; epistemic opacity; types of simulations in science; simulations as technological singularity; and detailed methodological discussions of numerous examples of the use of simulations in different sciences, from high-energy physics to climate science, exploring their features and limitations.

A smaller number of studies analyze of the ethical aspects of computer simulations, although an elaborate discussion of the ethical aspects of computing has been initiated (10) and developed through analysis of such pressing issues in computing as the ethics of big data. (11) Among the ethical and social issues raised in connection with simulations, the ethics of virtual reality, (12) the policy of simulations, and professional ethics of simulationists (13) receive priority. Issues immediately related to simulation ethics include decision making based on the simulations involving a model that misrepresents the target system or represents it without sufficient detail; the responsibility of code designers for the representational accuracy of the model and its implicit assumptions; the effect of the personal qualities and moral conduct of researchers on the credibility of simulations; an elaborate discussion of these topics can be found in Durán's study. (9)

However, in the course of their development and application, simulation codes not only develop but also cease to exist. The purpose of this paper is to explore the issue of intergenerational justice, arising from the loss of both the knowledge embodied in simulation codes and the knowledge and practical skills of using the codes in the community of simulation practitioners. Intergenerational justice (in technology) is referred to here as the equitable distribution of risks and burdens (related to the benefits of particular technologies) between the previous and next generations (see, for example, (14)). In section 2, we discuss the problem of the loss of scientific and technological knowledge through the example of nuclear technologies and simulations in nuclear and particle physics; section 3 deals with the disappearance of knowledge and codes in the field of nuclear interactions simulations, the structure of these codes, and the reasons for their discontinuation; section 4 analyzes the communicative and social structure of small groups of nuclear code developers and the ways in which these groups make decisions; in section 5, the model of four types of trading zones guides understanding of the structure and evolution of the actor networks developing simulation codes, as well as the degeneration of some codes into languages that supplant other codes; section 6 raises the problem of intergenerational justice, which attracts attention in the philosophy of technology and is key in the discussion of sustainability, in connection with the issue of preservation and transmission of simulation knowledge and codes;

section 7 proposes an answer to objections to the possible written explication of available knowledge about simulations, emphasizing the inherent nature of tacit knowledge in simulations and the role of practice in trading zones. In the conclusion, we have formulated an ethical requirement to support and develop simulation trading zones across generations to meet the requirement of sustainability of simulation practices and expertise.

1. **The problem of losing knowledge**

Very important questions of the preservation of critical technological knowledge arose in the studies of MacKenzie and Sims. (14–15) Sociologists who studied laboratories involved in the development of nuclear weapons found that a substantial portion of the knowledge about nuclear devices and technologies was embodied directly in technological products and their practical use, representing the tacit knowledge described by Polanyi. (16) It turned out that during the Cold War, knowledge of how to create nuclear devices was maintained in the community by virtue of the fact that the political environment and government programs continuously required the design and testing of new and emerging types of nuclear weapons. (15) However, with the end of the Cold War and the cessation of the production and use of new weapons, knowledge and technology began to be lost, and the threat of uninvention of nuclear weapons arose. Close in meaning to 'uninvention' is the concept of 'undone science,' (18) defined as 'areas of research that are left unfunded, incomplete, or generally ignored but that social movements or civil society organizations often identify as worthy of more research.' (p.444)

Recognizing the severity of the problem of nuclear knowledge protection, Sims (16) raised objections to MacKenzie, (15) premised on the experience of the Stockpile Stewardship program. The program preserves knowledge and experience by recognizing that technological knowledge is not static but continues to develop and evolve beyond the Cold War, and that perceptions of credibility of technology are intertwined with institutional changes and practices. The program, proposed to Congress and the Clinton administration in the 1990s (16), 'has focused on maintaining weapons expertise through a huge investment in non-nuclear experimental facilities and state-of-the-art computing and simulation capabilities.' (p.325) In this program, credibility of nuclear technology was no longer associated only with warheads, but also with computer codes, offered as the main link integrating various types of technologies. Simulations were the primary means to assess both the safety and performance of nuclear devices. The USDOE proposed that '[c]omputational simulation will be an essential (and sometimes the only) means of predicting the effects of materials aging on component and weapon performance.' (19) Thus, a significant part of the scientific community previously involved in the development of nuclear technologies shifted focus to the development and application of computer simulations, providing sustainability and institutional repair of knowledge and technologies.

Thus, computer codes used in simulations constitute a special type of technology, developed and distributed in ways different from other technologies. In particular, Sundberg proposed a convenient typology of simulation codes, (20) placing along the axes of 'use' and 'distribution' many codes found in meteorology, oceanography, and astrophysics. Codes, according to Sundberg, (20) can be for free or limited distribution, as well as for open or closed development. In our taxonomy, in the field of nuclear and particle physics, codes such as Geant4 (21) are free in distribution, developed by worldwide distributed collaborative teams. The support required for such codes is minimal, because such a collaboration is a self-developing and self-sustaining ecosystem, constantly replenished with new team members, which allows it to move in time as a social wave while preserving knowledge and transmitting it to future generations. In the presence of a sufficient number of high-energy physics or accelerator experiments of an open nature in the world, such simulation codes will continue to exist and develop without special efforts to protect them. We are not aware of open developer membership codes with closed application, because developers are usually the first (and sometimes the main) users of the code.

Codes with closed development (that is, created by a narrow circle of developers) are usually more specialized, designed for a more limited range of problems and phenomena. Among codes of closed development there are those of free distribution, such as FLUKA (22) and PHITS, (23) and closed distribution, such as MCNPX. (24) The existence of closed codes is conditioned both by a narrower range of tasks and applications and by the closeness of the field of application (as shown by Pelowitz (24)). Regardless of the mode of distribution to users, the specificity of closed development codes stems from the institutional need to maintain their existence, as well as developing, preserving, and transferring knowledge and technologies to new developers. Unlike open codes, which are open code ecosystems, closed codes can cease to exist in the absence of external institutional support, resulting in the loss of knowledge and expertise.

2. **Loss of simulation expertise**

The problem of the loss of simulation expertise—the so-called 'software rot'—has been known since at least the 1990s. As Galison put it (25), 'In more recent times this phenomenon has become known as 'program rot'—people and machines move on, leaving older programs dysfunctional, often irretrievably so.' (p.141) In a broader sense, the loss of functionality of computer programs can be conceived as a loss of knowledge of generations of physicists and programmers who have been developing programs, sometimes for decades.

And although there are myriad lost programs, even in the field of nuclear and accelerator physics, we will give two recent examples of codes that risk disappearing before our eyes. We will first consider the structure of such codes. The nuclear and accelerator simulation programs discussed above, in particular per Böhlen, (22) (FLUKA) Agostinelli, (21) (Geant4) Mokhov, (26, 27) (MARS15) Pelowitz, (24) (MCNPX) and Sato, (23) (PHITS) are structured as follows: to simulate the passage of high-energy particles produced by accelerators through the substance, the entire process is divided into two levels, micro- and macroscale. The macroscale—that is, the transportation of particle flows through macroscopic matter—is described by solving the well-known Boltzmann equation. However, on a microscale, when a description is required of the elementary interaction of an individual particle, such as a proton in a separate nucleus, scientists deploy rather complex computer routines called event generators. The development of generators is necessary because an exact theoretical and mathematical solution for such a problem is not known, and many approximate empirical and heuristic approaches are required; the creation of these is both a science and an art. Behind each such generator there is either an individual developer or a small group of developers.

Physics, as described by generators, is a conglomerate of diverse and sometimes heterogeneous models. Currently, there exists no complete microscopic theory of the atomic nucleus and nuclear reactions; instead, a multitude of semi-classical models describe the processes of interaction of high-energy particles with a nucleus. For instance, in order to explore how a highly energetic particle (like a proton) interacts with a metal block, a simulationist has to engender a model of a block embedding a multitude of computational procedures, (created by modelers) describing elementary processes (i.e., procedural knowledge of what such a proton can do with individual atomic nuclei it encounters on its path within the block).

Many process models describe the first step of the interaction as a so-called 'intranuclear cascade,' when an energetic particle strikes nucleons distributed inside a nucleus as a Fermi gas, similarly to billiard balls. Some nucleons then sprinkle out of a nucleus this way. At this stage, the nucleus is represented as a quantum mechanical gas of nucleons. However, after that, other nucleons that acquired energy in the cascade can 'evaporate' out of the nuclear remnant; at this stage, the nucleus is a rotating drop of a hot quantum liquid. Sometimes an intermediate model, such as the preequilibrium model, is included between these two stages; this often improves their agreement with data, though not in all cases. Thus, the nucleus during the interaction is represented by two or three different large models that bear upon quantum mechanics.

Even those two or three basic models can manifest in many different ways; a multitude of models will increase the overall generality. For instance, when describing how a nucleon leaves the excited nucleus, it is necessary to model the surface of the nucleus, and there are different approaches to modeling that surface and its shape. Also, when a stricken nucleon is moving within a nucleus, it is important to model how the nuclear density changes from the center to the periphery. There are several ways to describe that change, such as splitting the nucleus into several concentric zones with different densities in each. The emission of individual nucleons or groups of nucleons can be modeled not only as the evaporation of one or several nucleons from the surface of the liquid drop but also as the asymmetric fission of the excited nucleus into the nucleon or the cluster of nucleons and the remnant. A discussion of the epistemological implications of combining different models in codes can be found, for example, in Morrison (8) and Pronskikh. (28)

Akin to an alchemist, a generator developer works with a team for decades, often for his or her entire career, at reconciling various code ingredients to achieve empirical adequacy (validation) and, therefore, applicability to new scientific problems. The fact that the number of authors of a generator usually ranges from one to three people is indicated, in particular, by Merz, (29) citing examples of such generators used in high-energy physics in large scientific laboratories like CERN or PYTHIA. (30) Popular generators of the present or recent past include those of Bertini, (31) Boudard, (32) and Mashnik, (33) included as components of transport codes (that is, frameworks that combine micro- and macroscales), like those of Agostinelli, (21) (Geant4) Mokhov, (26, 27) (MARS15), and Pelowitz, (24) (MCNPX). One of the most popular, the model of Bertini, (31) has practically stopped development with the departure of the author and the main developer, and is being included in the current transport codes in the same form as in the past. Indeed, the same fate awaits the CEM and LAQGSM generators, (33) whose developers passed away recently and, in the author's view, did not leave behind them groups capable of continuing to develop the code independently. Somewhat earlier, from the mid-2000s, the transport code CASCADE, (34) which comprised both the solver of the Boltzmann equation and the original generator of nuclear-nuclear collisions, disappeared into oblivion with the demise of its author. All these missing or disappearing codes are examples of closed codes, or codes developed by narrowly closed groups or individuals affiliated with specific research laboratories.

3. 'Alvarez' codes

This sad list of extinct and 'rotting' closed codes is extensive. However, it seems necessary to try to figure out the reasons for their cessation, in addition to the obvious funding problems. All these codes were developed either by an individual or by a narrow group under the leadership of an individual who possessed knowledge as well as personal direct daily microcontrol of all stages of code development, exemplified in the structure and interactions in the Alvarez group (35). Most of the comments, descriptions, and instructions were understandable to the group leader (or few group members); and a significant part of the important information and code work was either not committed to paper at all, or was designed for the leader's personal use or that of his or her authorized representatives. Given that the codes number hundreds of thousands of lines, it seems necessary to have at least a comparable amount in an accessible form description to make the code comprehensible to an external reader. The developer-leader of the group makes all decisions.

The head possesses the skills and knowledge necessary to work on and understand the functioning of each of the sections of the project, and he or she even sometimes, by choice, replaces one of the group participants to play the role of a subject specialist. All tasks for the group members come from the single project manager, who also finances the project. The head is responsible for the adoption of all decisions regarding the code, from the use of specific programming languages to the selection of results for publications and reporting. This method of organization in science is most natural for groups of students and postdocs led by a professor, and so serves as a traditional way of organizing small and medium-sized research projects. Although discussions regarding both the technical aspects and the formulation of knowledge claims (analysis and interpretation of modeling results) in such collaborations do not exclude the participation of several group members, the group's opinion is usually ultimately determined by (or reduced to) the views of the group leader—the phenomenon called 'groupthink' in social epistemology. (36) A shared intention to achieve the epistemic goals of the experiment may be either present among the group members (the leader and his subordinates) or absent (external specialists, such as system administrators performing certain functions). Thus, the small size, the rigid hierarchy, the manager's knowledge of all stages of the experiment, and the groupthink when making decisions are the main distinguishing features of this way of organizing code developers. Although scientific groups in which the opinion of the authority usually determines the group opinion are sometimes called Alvarez-groups (37), (36) attributes such reductions of group opinion to groupthink, of which she wrote that 'It occurs when a group of individuals aims to reach consensus on a controversial topic. Peer pressure, as well as pressure from those in authority (if present within the group), leads dissenting individuals to change their minds and, perhaps as important, not to share their knowledge of contrary evidence. ' (p.31) One can say that this method requires joint intentions but does not require the emergence of a collective epistemic subject. The interaction between the participants is mainly cooperative rather than collaborative (as is the case with open codes, such as that of Agostinelli (21)).

4. **Simulation codes as trading zones**

To examine how simulation code can and should evolve, it seems interesting to consider the simulation as a trading zone both between theory and experiment and between developers and users. (38) As noted, being in the center of the trading zone, the community of simulation experts found themselves in the 'pariah' position[3] in relation to theorists (who always hold the highest epistemic status) as well as to experimenters, because (25) 'the simulators became indispensable as links between high theory and the gritty details of beam physics and particle collisions. But just as they occupied an essential role in this delocalized trading zone, they also found themselves marginalized at both the experimental and the theoretical end of particle physics.' (p.155) Users and computer programmers are sometimes labeled 'inhabitant,' (39–40) and it can be difficult to transfer knowledge between the 'inhabitants' of different generations, as when using certain programming languages or techniques. To keep zones inhabitable in our understanding means that it is necessary to support and maintain the existence of a community of users and code developers (i.e., the life of the trading zone).

In particular, Gabriel (40) writes, 'languages that encourage abstraction lead to less habitable software, because its expected inhabitants—average programmers working on code years after the original designers have disappeared—are not easily able to grasp, modify, and grow the abstraction-laden code they must work on.' (p.20) Here, an understanding of the code used implicitly is close to the trading zone, which, in addition to inanimate actors like technologies and locations, encompasses human actors and the relationships among them existing in such heterogeneous networks. Such human actors in trading zones inhabit the codes, which we consider non-human network actors (here we use a definition of networks close in meaning to Latour (41) and assimilate trading zones to the states of such actor networks).

### 4.1 Enforces zone codes

The most striking example of *enforced zones*, in the view of Gorman, (38) may be the interaction between Romans and rowers in the Roman galleys; the rowers, significantly different culturally, did their job for water and food under the threat of physical punishment. Implementing this type of exchange required only an alternation of encouragement and punishment. Close cultural differences and mechanisms of interaction can serve as relations between scientific communities, as well as any communication situation in which one group sees no usefulness for its language beyond the bounds of its own discipline (e.g., theoretical physicists in relation to experimentalists). The second (subordinate) group may also be reluctant to learn the language of the dominant group and integrate it into their language. Such trading zones appear first in the model. In science, the first type is similar to the initial stage of an interdisciplinary collaboration: to start developing simulation code, scientific centers must bring together physicists, programmers, and application scientists who all use different professional languages, forcing them to collaborate using institutional mechanisms. If successful, such coercive but heterogeneous cooperation may result in the creation of a simulation code. In an institutional context, a coercive trading zone can arise when a group is organized to develop simulation code, for which institutional incentives are used.

---

[3] An apt metaphor used in Galison. [25]

A zone of this type is the initial stage in the evolution of any simulation code, the creation of either a collaboration or a group of code developers at a scientific center or laboratory.

**4.2 Fractionated zone codes**

If the collaboration is successful, in the next step, according to the generalized model of exchange zones, the zone can become *fractionated* but remain heterogeneous. The two main traits of such a zone are first, a changing topology, because specialists originally in different departments move to one laboratory (or a few neighboring ones), where their interaction becomes more intense; and second, the development of common interests that supersede institutional coercion, particularly the development of a goal shared by all participants to create a specific technology (or to obtain certain knowledge) through cooperation or collaboration. However, the disciplinary languages at this stage remain different. Therefore, two main mechanisms of interaction between disciplinary cultures can be implemented.

The first method of interaction associated with material culture is embedded in the concept of so-called boundary objects. Boundary objects (42) are objects with different meanings in different cultures, but local coordination of their actions regarding the treatment of these objects is possible. In the context of an interdisciplinary scientific project, products and technologies developed by multidisciplinary teams, such as simulation codes, can act as boundary objects (38). They serve as boundary objects between developers of different units of code, and for the exchange between developers and users (representatives of specific sciences and technical disciplines) of this code. In particular, emphasizing the versatility of both applications and roles in which simulation codes as boundary objects are played, Merz (29) notes that codes are simultaneously epistemic things (43) for their developers and technological 'black boxes' for users (such as experimental physicists).

At the beginning of a joint simulation project, the groups of specialists involved only coordinate their activities locally, in connection with the joint creation or use of some code-boundary object. However, as the interaction deepens, they partially acquire *interactive expertise* in each other's languages, as with nuclear physicists or engineers and computer programmers in our example. Interactive expertise serves as the second mechanism of community interaction in fractionated exchange zones (of the second type), and can be defined as a partial mastery of the professional language of another community (e.g., physics or programming languages) at a level sufficient for full-fledged communication in the new language within certain contexts (38). Collins (44) refers to the ability described by sociologists who studied the community of gravitational physicists to answer questions about gravitational waves, also compiled by gravitational physicists, in such a way that representatives of the gravitational community could not distinguish the answers given by the sociologist from the responses of physicists. In the case of interdisciplinary projects such as particle transport simulation codes, whose development teams comprise scientists of different specialties as well as users with different specializations, communication in the second type of zone can begin with the acquisition of interactive expertise in each other's languages. As the project develops, in our opinion, the lead becomes an expert in the creation of jargons (special terms understandable to all participants), then pidgins (simplified languages from a mixture of elements of two active languages), and then full creoles (comprehensive scientific languages), transforming the zone into the third type.

Simulation codes' development into creoles can be facilitated by the minimized use of original physical theories in such an exchange. Instead, a user is offered certain narrative interpretation, and the language in which developers communicate with users will contain many simplified model terms. As a result of these processes, the culture gradually becomes more homogeneous as all participants in the exchange (i.e., users and developers) learn the same simplified languages, and the zone converts into the third type. At the same time, even using the same simulation approaches for different tasks, the exchange participants remain within their scientific specialties (25): 'A chemical engineer could speak easily to a nuclear physicist in this shared arena of simulation; the diffusion equation for chemical mixing and the Schroedinger equation were, for purposes of this discussion, practically indistinguishable. Yet this is not to say that the nuclear physicist and the chemical engineer themselves became identical or interchangeable—far from it.' (p.152) All codes that have developed for some considerable time and found their users have reached this stage of their development—in particular, all the nuclear and accelerator codes discussed in this paper, both transport codes and generators.

### 4.3 Interlanguage zone codes

The next, third type of trading zone (38) is called *interlanguage*. The nature of the third type of trading zone is manifested in the fact that as a result of joint practice, accompanied by the development of interactive expertise, a new simulation tradition develops; this tradition can be likened to a full-fledged creole arising from the interaction of two or more cultural disciplines. Such analogues as nuclear technology, biochemistry, and phenomenology of elementary particles inform the way we view interlanguage trading zones. Such codes become languages in their own right, universally used. In our view, good examples of codes that have reached the creole level are the codes of Böhlen, (22) (FLUKA) Mokhov, (26, 27) (MARS15) Pelowitz, (24) (MCNPX) and Sato, (23) (PHITS). When such codes become sufficiently widespread and inclusive, they gradually turn into stable research programs; that is, they serve as the official codes of scientific centers, laboratories, and communities. University students and users then learn to use such codes from the student bench; lectures, courses and schools are organized, requirements for using such programs are formulated for scientific experiment projects, and codes become the official standards for centers and laboratories. It could be argued that between the second and the third type of trading zone, the code changes its role and turns from an epistemic object (in Merz's terms (29)), studied and created by developers, into a technological object, or a tool for solving cognitive tasks in other user communities.

Unlike generators, (31–33) wherein the decision-making role is assigned to the sole leader, teams for codes such as those of Böhlen (22) and Pelowitz (24) may have a more complex structure. Groups of their developers can be quite extensive, so much so that management can be carried out by boards, more reminiscent of collaboration in experimental physics. However, this analogy is incomplete, because such groups maintain a strict hierarchy of decision making with formally defined leaders. Therefore, the method of making a group decision can, as in the case of the Alvarez code, be reduced to groupthink, or opinions of dominant participants (or indeed a single participant). This way of organizing can be likened to big science organizations such as the Manhattan Project. If the clear leaders of such projects leave or depart, then the project itself may come to an end, and the code may 'rot.'

### 4.4 Subversive zone codes

Having existed for some time at the same level of quality, the trading zone can evolve to the fourth type. This type of exchange zone in the generalized model (38) is also called *subversive* by the authors. At this stage, new participants from other disciplines who join the new culture can no longer introduce new elements to it, because the new discipline, which has now turned into a 'normal science' in the Kuhnian sense, imposes ready-made models on the 'converts.' All new ideas coming from other areas are suppressed by the dominant paradigm of the new culture—that is, by the norms and values already circulating in the zone. One example of this type of zone would be Einstein's Special Theory of Relativity (SR), which replaced Newtonian mechanics. In the majority of applications, the SR has certainly become the dominant theoretical framework, and specialists who resort to mechanical calculations concerning rapidly moving objects must accept the SR language without any modifications and cannot make any innovations or changes to its form without risking rejection by the community. The quark model served as the same type of subversive zone in physics, displacing all other models of strong interactions (e.g., the Regge theory).

The code of Agostinelli, (21) in our opinion, has achieved the level of a trading zone that sets the standards of thinking in a whole field of high-energy physics. By contrast with closed codes that have reached the level of development of the third type of zone, the code of Agostinelli (21)—an open code developed by a collaboration of physicists and programmers from all over the world—is organized in a different way from Alvarez codes. Collaboration with open membership breaks up into a core and periphery in a way somewhat similar to big science. However, there are several structural differences. Firstly, the communicative core divides into several boards ('institutional,' 'executive,' etc.) and committees (for speakers, publications, etc.), whose members, upon completion of their terms of service, rotate to other bodies within the core or are replaced by other members of the collaboration. All decisions, by contrast with the previously discussed groups of Alvarez codes, are deliberate, not the result of groupthink. Secondly, the entire core does not interact with the peripheral members of the collaboration, but rather with the direction leaders within the core, who coordinate the work of a certain branch of the collaboration responsible for a specific area. In this sense, there is a core of collaboration (similar to the mega-science experiment collaboration cores) with a communicative way of organizing and a communicative rationality, but the division of the peripheral parts recalls the organization of big science. The core members have interactive expertise in aspects of the experiment beyond the limits of narrow specialization, while the periphery contributes expertise in its members' technical fields.

The general intention to create a holistic multifunctional code and to maintain its functioning as a central boundary object is shared by all members of the collaboration, distinguishing these methods of organization from those of big science and Alvarez groups, in which a commonality of intentions was not required the same extent. Participants in this code move, as discussed above, like a wave, presenting and uniting different generations of developers, constantly replenishing their numbers with new members as previous members leave the group. The code no longer just remains a center or laboratory code, but a *lingua franca* mastered at the training stage by all students of a given specialty. Moreover, such a code crowds out the other codes in the field of its application and related fields, as noted above. The supplanting carries a certain risk of neglecting systematic errors, given that simulations of even the same phenomena, when implemented in varying ways (that is, with independent codes), allow users to identify an error made in any one of the codes. It is the displacement of the fourth type of zone code, an emergence of a single all-encompassing code, that is one of the reasons simulation is gradually at risk of becoming a technological singularity, (45) because knowledge embodied in codes supplanted by dominant codes of the fourth trading zone may be lost for future generations. It is worth considering why the preservation and transfer of technology for descendants, such as simulation codes and the knowledge embedded in them, is a requirement of justice.

## 5. Sustainability and intergenerational justice

The problem of intergenerational justice is directly related to the preservation and transfer of technology and technological knowledge for future generations (14). The Pure Intergenerational Problem (PIP) was originally formulated (46) around the example of pollution. Gardiner (46) framed the problem as analogous with the well-known philosophical problem of the Prisoner's Dilemma, wherein two statements—each logical on its own—lead to a contradiction. If, from the point of view of collective rationality, it makes sense for different generations to cooperate, limiting the pollution of each and thereby improving the outcome, also for each, then from the point of view of individual rationality, a more favorable model of behavior is the rejection of cooperative behavior and overpollution in each individual generation. Projecting PIP on a situation in simulation with program codes and the knowledge embodied in them, the following statements arise:

1. From a collective point of view, it is more beneficial for each generation for all generations to preserve and transmit the codes they created for the next generation, because this maximizes the useful outcome for each.

2. From an individual point of view, however, when a given generation has sufficient power to decide whether to keep software products or not, it is rational for that generation to refuse to expend resources on preservation and transmission.

Along with these statements, a number of other contradictions apply not only to simulations and computer codes that implement them but also to technologies in general. As noted by NAPA, (47) 'no generation should (needlessly) deprive its successor of the opportunity to enjoy a quality of life similar to its own.' (p.7) The first priority for NAPA is this to eliminate the risk that affects both current and future generations, while phenomena that pose a risk to only one generation have a lower priority. Phenomena closer to the current generation are more likely to share problems with that generation than with the distant generations. In addition, little can be said about the technologies and knowledge distant generations will possess. Computer codes that simulate processes in modern accelerators and high-energy physics (for example, all the generators for collisions of particles and nuclei described above and transport codes based on them) are also likely to adequately describe processes in accelerators of the near future under linear development of technologies (that is, normal science in the Kuhnian sense). However, in the distant future, if we anticipate a revolutionary change in accelerator or experimental technologies, current simulation codes may turn out to be nonusable, just as the transfer of technology for lighting rooms with candles retained its significance only until the invention of whale oil lanterns, gas lighting, and then the electric incandescent lamp. In other words, the greatest responsibility lies on the scientific community with the transfer of simulation technologies to the closest generations of future researchers.

A separate question in relation to the existence of simulation codes is what can be considered a generation. Regarding closed development codes, the tenure duration of the leading developers can be considered the duration of a generation, which can be 30–50 years. In relation to open development codes, it is still difficult to define this concept more precisely, because less than 30 years have passed since the beginning of their development, and it will be possible to determine when a generation ends only when some open source codes cease to exist or when they change their character in a fundamental way. Thus, it seems that at the given moment, a generation in simulation codes can be considered a period similar to a demographic generation.

Simulations have created one of the most rapidly developing areas of technology in recent decades. To predict the problems and solutions they will bring to this area of intergenerational ethics, it seems necessary to turn to the experience of other technologies, including those employing simulations, such as nuclear technologies. One striking example of the inverse kind is the technology of nuclear waste treatment (48). In the event that the current generation accumulates nuclear waste, the burden of processing it rests with future generations; if the current generation does not spend money on its disposal, then the well-being of future generations may suffer. In the case of computer simulations considered in this paper, given that the current generation is not assuming the burdens of labor and expense to preserve the knowledge embodied in computer codes, future generations will have to reinvent many of the methods currently used. Thus, both in the case of simulations and in the management of nuclear waste, the problem can be reduced to the PIP discussed above. In both cases, the well-being of future generations depends to a large extent on the efforts of the current generation.

Further in the discussion of PIP, researchers on the ethics of nuclear technology turn to Barry, (49) who formulated two key ethical principles on which to base approaches to addressing PIP and the regulatory aspects of renewable technology development: the principle of responsibility and the principle of vital interests. According to the first, a problem in the occurrence of which someone is not guilty requires *prima facie* compensation, to the extent that the contrary has not been proved. In accordance with the second principle, a person's specific position in time and space—that is, belonging to a certain generation—cannot serve as a basis for violating that person's vital interests; the interests of the people of the future have the same priority and importance as those of the people of the present. Compensation for future generations of simulation researchers will not be required if the computer technologies developed so far for the performance of simulations are transferred to them without the loss of important information and essential skills which would prevent their use. If such losses occur, though, descendants would have to recreate and reinvent the methods of simulation in order to support the safe use (and mitigate the attendant burdens) of the technologies they will inherit from the present generation.

This problem is aggravated by the fact that computer programs for simulations are currently extensive frameworks fleshed out for decades by the efforts of hundreds of top-class specialists, and the development itself costs a considerable sum.

An implicit prerequisite for the normative requirement of the preservation of knowledge embedded in simulation codes is the assumption that the transfer of such knowledge to future generations is always good. Because most of the codes discussed are related to modeling the processes of nuclear interactions, one could argue that maintaining the knowledge of nuclear technologies is not always a boon because it can potentially find dual use in nuclear weapons. There are two objections to this. First, as was recognized by (18), the cessation of nuclear research and practice can lead to the uninvention of these technologies, as discussed in the paper. Despite the idea that these technologies themselves do not always find peaceful use, it can be assumed that in the context of an insufficiently globalized world and the possibility of military confrontation between military blocs or individual countries, the uninvention of nuclear technology by one of the parties to the confrontation, while the technology is preserved by others, can weaken one of the parties and even provoke a military conflict. Thus, the danger of the uninvention of nuclear technology lies in the asymmetry that it triggers and the difficulty of monitoring the equal observance by all parties of the conditions of uninvention. One of the regulatory requirements for preserving nuclear technologies and transferring them to future generations is rooted in this danger. Second, and methodologically, the simulation requires a developed computer infrastructure along with extensive all-around support, and therefore it requires the sort of calculations that can be controlled by an expert community. Also, the expert community will know the potential of the simulation codes, so the distribution of the data necessary for the calculations can be limited to those data that allow only peaceful use. For example, data on the sensitive isotopes can and usually are excluded from libraries of cross sections for nuclear reactions intended for open distribution. Sustainability additionally emphasizes that the community should be regularly updated and allowed to share knowledge (i.e., allowed to function as evolving simulation trading zones). The most self-sustaining are open codes due to their reliance on diverse funding sources as well as the broad range of participatory motivations.

The need to maintain simulation trading zones logically follows from the requirement to preserve and transfer procedural knowledge enclosed in simulation codes because it is in the form of evolving trading zones, as shown in this paper, that codes are developed and expertise is transferred. Trading zones associated with open development codes turn out to be the most sustainable, so another related requirement is to create conditions for the evolution of all possible closed trading zones to the level of open zones. The latter, according to our application of the model (38), constitutes one of the final stages in the evolution of trading zones. In the case of the first two types of zones, enforced and fractionated, which we associate with closed codes, one cannot talk about the common language of the professional participants, but instead one can discuss such mechanisms as boundary objects (codes) and interactional expertise. In the last two types, interlanguage and subversive, code is a kind of language that all participants can use (for example, participants can analyze physics problems in terms of the Geant4 code procedures). Such a code-language is simultaneously algorithmic, mathematical, nuclear-physical, and accelerator-physical; that is, it is a synthesis of several professional languages. At the same time, such a language is a sequential stage in the evolution of interactional expertise. In most cases, in our opinion, such an evolution is possible and needs to be supported. However, in certain cases, for example, when closed codes, in view of their special scope of application, cannot be made open to every stage of evolution, it is necessary to provide them targeted support so that open codes do not displace them from their ecological niches (i.e., they must be kept inhabitable).

### 6. The role of tacit knowledge in simulations

A number of objections can be raised to the proposal that logically follows from the previous analysis premised on PIP, namely that there is a moral imperative to keep the trading zone arising during the development of simulation code for several generations (i.e., until the technologies change in a revolutionary fashion). Because descriptions and manuals are created during the development of the code to preserve information about the physical and mathematical principles underlying the code, future generations with access to such written sources could be able to restore knowledge and resume previously discontinued code practices, even in the absence of an appropriate trading zone. Thus, to preserve the knowledge contained in simulation codes, it would be sufficient to simply document them in detail. To rebut such an objection, it is necessary to consider the tacit knowledge emerging and transmitted in simulation trading zones. One frequently noted problem is the lack and narrowness of such documentation. (29) However, there are problems of a fundamental nature that impede the reduction of the problem of intergenerational transfer of knowledge to the transfer of written documentation. Although the concept of tacit knowledge can carry several meanings, in particular, MacKenzie (15) believe that nuclear technology specialists can acquire tacit knowledge by studying records. It is worth noting (50) that tacit knowledge acquired during practice is characterized by the fact that the carriers of such knowledge may be ignorant that they possess it. Tacit knowledge is thus inexplicable in writing.

Collins (51) was one of the first to draw attention to the role of tacit knowledge and tacit presuppositions in scientific practice, while Soler (52) points to the connection of tacit knowledge of experimenters in science with the opacity of experimental practices. Premised on her analysis, Soler (52) challenges the substitutability of experimentalists in view of the fact that experimental practices are opaque with respect to description and justification. Because individual opinions of scientists regarding experimental results are predicated on their trust in other participants with tacit knowledge of specific issues—that is, on their epistemic dependency, the degree of which largely varies contextually and historically—there is no sufficient reason to consider such experimental facts nonhistorical. But there is another important aspect that we can identify in (52): narrow groups of scientists create knowledge, and each group member has tacit knowledge inherent only to him or her, which is why the group members complement each other in the course of epistemic practice. Group members are literally indispensable to each other.

The role of opacity in the practice of simulations has been discussed in detail. (53-54) It follows (53) that simulation codes are as extensive as, for example, particle transport codes and their generators, and the calculation requires so many steps and loops (e.g., as when calculating by the Monte Carlo method) that it is usually impossible to trace each step of code execution up to each executed instruction. For this reason, users have to rely on both experimental strategies, becoming partly experimentalists and epistemically dependent on developers and other users. Epistemic dependency on developers, as well as nonuniversality of documentation, is elicited by the opacity of simulation codes noted above. Users have to rely on the tacit knowledge of developers, which underlies the creation of codes; and being tacit, such knowledge, by definition, cannot be completely transferred to paper.

Another manifestation of tacit knowledge is the ability of code developers to understand code which, while growing in size, loses much of its logical structure, turning into a (55) 'big ball of mud.' (p.12) For discussion, see (56). Although organizationally, especially initially, simulation codes are not structureless systems, they certainly undergo structural erosion during development, which bears upon piecemeal growth of codes, especially in distributed systems created simultaneously by many developers, leading to the loss of rational and modular organization. Codes that were originally created for temporary use, or to solve small and specific problems, keep growing to enormous size, losing their original structure. The only way for the developer to keep the ability to maintain code is to rely on tacit knowledge. One sign of code structure erosion is the introduction of kludges—that is, parts of the code that are unprincipled or not sufficiently substantiated (56–57). Such kludges are usually undocumented and introduced temporarily in order to overcome temporal technical difficulties in the code and its use, but then the reasons for their introduction are forgotten and kludges remain in the code for the entire duration of its existence. Either the mechanisms of functioning of kludges remain clear to the developers who were involved in the origin of the code (and then such knowledge can be explicated, but this is rarely done, because the rationale for such kludges is often not sufficiently scientifically sound), or the developers acquire a verbally inexpressible feeling for using a kludged code to achieve their research goals. In the latter case, a novice developer of the next generation joining the developer group gains knowledge about the effective use of the similarly kludged 'ball of mud' through cautionary tales passed through the community and in the course of experimenting with the code. Because learning simulations in the relevant community involves experimenting with code, it helps to form tacit knowledge.

Yet another obvious problem of simulation codes—trading zones stopping in their development on the third type, that is, closed (Alvarez-type) codes (including all transport code generators and most transport codes in general)—is that because a strategy for developing a code (which consequently quickly gains an eroded structure) is determined by the leader of the group, knowledge of the reasons for introducing one or more kludges remains exclusively their prerogative (or in rare cases, that of two or three people). This observation highlights the role of the structure of the communities that comprise the trading zone, emphasizing the role of communication and the transfer of knowledge and skills during the evolution of trading zones in socializing in the community and learning through practice. This, in turn, requires the institutional maintenance of the existence of such a trading zone across generations, as well as the gradual inclusion of new generations in the learning process to transmit tacit knowledge as a form of social relay.

In simulations, tacit knowledge can be understood as case law in the sense of Polanyi, that is, the reuse of solutions whose application in the past led to success. However, there are drawbacks and limitations to applying this approach to simulations. For each code, the number of such tacit solutions that a practitioner has learned over the course of a career is large. This multitude includes an understanding of which set of tens and hundreds of input parameters, of data transferred inside the executable module or structural module connected to the work, will provide the most accurate and adequate solution for each of the hundreds of possible applications that deploy simulation code. The quick finding of the right solution and its competent application is a skill honed over years of activity and varies in different domains of code application. The mere study of ample manuals alone is much less effective than communication among professionals and inclusion in the community, which significantly reduces the time required to train a professional. This can be assimilated to the training of a virtuoso musician: despite the fact that music notation and descriptions of performing techniques have existed for centuries, training a music performer requires mentoring and immersion in the complex life of the community (one can read about the difficulties of preparing virtuoso musicians, for example, in (58)). Because simulations can be considered a way of applying experimental research strategies, that is, because they are the cheapest way to experiment, it is challenging to propose any other alternative that would be an equally effective method of preserving simulation knowledge because simulations are for traditional experiments. A distant analogue seems to be thought experiments; however, they cannot serve to fully replace simulations because simulations are distinguished by extremely high technical complexity, which leads to their epistemic opacity, which is absent in thought experiments. In this sense, simulations are a unique way of obtaining knowledge.

## 7. Conclusions and recommendations

In this paper, we examined the challenges posed by the loss of knowledge embodied in simulation codes, code extinction, and their relationship to intergenerational ethics and PIP. The loss of technological knowledge, especially in areas such as nuclear technology, is a well-known phenomenon, and one of the solutions proposed by technology researchers was the transition of the expert community to simulations in order to preserve and repair knowledge. It turned out that simulation codes are also prone to rot and extinction. However, the so-called closed codes—that is, codes created and maintained by research centers and small groups of individuals—were typically shorter-lived. Such codes are developed by individuals or small groups rigidly and hierarchically led by such individuals, which we call Alvarez groups; such codes stop developing after the leader leaves or departs, because the decision-making mechanism in such groups is close to groupthink. We analyze simulations as a trading zone, which we understand as the state of a network of actors evolving over time, including developers, users, and non-human actors—that is, simulation codes that play the role of boundary objects.

Our analysis indicates that the closed codes developed in trading zones developing (in accordance with the generalized model, we distinguish four types of zones) stop their evolution at the third type, becoming creoles of scientific centers and laboratories, then gradually vanish. By contrast, the reign of open codes created by globally distributed collaborations with open membership and without defined boundaries can evolve to the fourth type, becoming universal languages that push other codes out of their ecological niches. Thus, from the point of view of preserving the knowledge implemented in codes, the risk for closed codes is their disappearance, potentially exacerbated by uncontrolled growth of open codes. The life cycles of the respective community networks constrain the existence of codes, because the development of codes relies heavily on tacit knowledge.

We argue that the cessation of the development and existence of simulation codes and the subsequent loss of the knowledge incorporated in them is an ethical problem, because—as with other critical technologies—the failure to preserve technologies developed by the current generation and the inability to transfer them to future generations will require these generations to rediscover the technologies to tackle inherited technological issues, thereby violating their vital interests. We suggest responsible and sustainable maintenance of simulation trading zones to prevent both the disappearance and mutual crowding out of codes, and to transfer knowledge as a social relay to future generations until radically changed technologies eliminate the importance of specific codes.

**Acknowledgements**

The author is indebted to Dr. N.V. Mokhov for fruitful comments and discussions. He recalls with appreciation the fascinating exchanges and collaboration with departed colleagues and prominent scientists, Drs. N.S. Amelin, V.S. Barashenkov, K.K. Gudima, S.G. Mashnik, R.E. Prael, and S.I. Striganov, whose expertise and role in the development of Monte Carlo simulation codes in elementary particle physics and nuclear physics significantly influenced his grasp of this field as well as deepened his immersion in the simulation community and its nourishing culture. He would like to thank two anonymous reviewers for their careful reading of the manuscript and valuable suggestions.

Fermi National Accelerator Laboratory is operated by the Fermi Research Alliance, LLC under Contract No. DE-AC02-07CH11359 with the U.S. Department of Energy, Office of Science, Office of High Energy Physics.